# Orientational Ordering and Binding in Alkali doped $C_{60}$ solids

K. Ranjan, Sarbpreet Singh, K. Dharamvir and V.K. Jindal[1]

Department of Physics, Panjab University Chandigarh- 160014

The binding energy of $A_3C_{60}$, a conductor, is described well by an ionic solid type calculation. This succeeds because there is little overlap between molecular wave functions on neighbouring sites, so that electrons are practically localized on-shell. This leads one to believe that even in $A_4C_{60}$ and $A_6C_{60}$ systems such calculation may suffice. However the on shell Coulomb repulsion is large for the $C_{60}$ molecule. So, for large charge on the anion, there is a possibility for some electrons to delocalize and go into the s-band. In the calculation of binding energy, we keep these delocalised electrons x, as a parameter and minimize the energy w.r.t. it. We take the intermolecular interaction to be arising out of a C-C potential of 6-exp form (Kitaigorodsky) and a screened Coulomb interaction between the anions and cations and among themselves. The screening is provided by the electrons delocalised from the anion which supposedly go into the s-band of the cations, and are modeled by a free electron fermi gas. The energy of the anion (to be added to the lattice sum) takes into account the on-site Coulomb energy, and is thus a quadratic function of anion charge. The delocalised electrons go into s-band whose position is estimated and corresponding energy added. Model calculations are presented for $K_1C_{60}$, $K_3C_{60}$, $K_4C_{60}$ and $K_6C_{60}$ for which the minimum energy state shows no delocalisation. Cohesive Energy dependence on Lattice constant is used to calculate Bulk Modulus for all systems. We have got a reasonably good resemblance with experimental values. Further, we observe that the cohesive energy shows poor resemblance with experimental values. This can be explained by invoking orientation in these calculations. Further, delocalisation of a fraction of electron at the centre of double bond show considerable increase in cohesive energy.

## INTRODUCTION

Carbon is a remarkable element showing a variety of stable forms ranging from 3D semiconducting diamond to 2D semimetallic graphite to 1D conducting and semiconducting nanotubes to 0D fullerenes. The closed cage nearly spherical molecule $C_{60}$ and other related fullerene molecules have attracted a great deal of interest in recent years because of their unique structures and properties.

---

[1] E-mail: jindal@panjabuniv.chd.nic.in

The $C_{60}$ molecule is a regular truncated icosahedron. The 60 carbon atoms in $C_{60}$ are located at the vertices of a truncated icosahedron where all carbon sites are equivalent. A regular truncated icosahedron has 90 edges of equal length, 60 equivalent vertices, 20 hexagonal faces and 12 additional pentagonal faces to form a closed shell[2]. The boundary between a hexagon and its neighbouring pentagon is a single C-C bond and that between two hexagons is a double bond. Because of the large size of fullerene molecules as compared to alkali metal dopant atoms or ions, the interstitial cavities in a $C_{60}$ lattice can accommodate various guest species.

The free $C_{60}$ molecule in its ground state electronic configuration has a completely filled HOMO level. The LUMO is split into two three -fold degenerate levels which get filled as the Bucky-ball is charged with more and more electrons. When the pure $C_{60}$ solid is doped with alkali metal (exohedral doping) $A_nC_{60}$ compounds are formed, and it has been observed that n can go up to 6 presumably filling the above described levels[1].

The most widely studied of these is $A_3C_{60}$ which is not an ionic solid, but a semiconductor. However, bands derived from LUMO levels are so narrow (which is because the $C_{60}$ molecules are sufficiently far apart in the solid), that an ionic-solid-like calculation gives the correct cohesive energy[2]. Hence even when $n > 3$ in $A_nC_{60}$, it is the molecular levels which are further filled, and binding is believed to be ionic -like. We have studied Binding Energy invoking a free rotation model; and then including orientations. It has been observed that the Potential Energy between two $C_{60}$ molecules becomes independent of orientation for centre to centre distance greater than about 9Å (Fig2); whereas orientation is important for alkali-$C_{60}$ interaction due to much smaller distances involved (Fig 3 and Fig 4).

## THEORETICAL MODEL AND CALCULATIONS

### General Interaction Model

The intermolecular potential energy $U_{l\kappa,l'\kappa'}$ between two $C_{60}$ molecules, identified by the $\kappa$th molecule in unit cell index **l** and $\kappa$'th molecule in cell **l**', can be written as a pair-wise sum of C-atom-atom potentials (C-C)of carbon atoms on these two molecules, i.e.

$$U_{l\kappa,l'\kappa'} = \sum_{ij} V(r_{ij}) \tag{1}$$

where the sum in Eq.(1) includes all the 60 atoms in each of the molecules. V(r) is the C-C potential for two C atoms situated at r distance apart, and is assumed to be of the form

$$V(r) = -A/r^6 + B\exp(-\alpha r) \tag{2}$$

The interaction parameters[3] A, B and $\alpha$ have been obtained by Kitaigorodski . The interaction parameters for C-C, C-A and A-A interactions are listed below

**Table I**

| Atom-Atom | A | B | α |
|---|---|---|---|
| C-C | 358 | 42000 | 3.58 |
| C-K | 171 | 39105 | 4.51 |
| K-K | 235 | 25784 | 3.4 |

### Free-Rotation Model

For the case of freely rotating molecules, i.e. at temperature high enough for orientational correlation to vanish, the $C_{60}$ molecule can be replaced by a spherical shell[7], with 60 carbon atoms uniformly smeared over the shells. We can then use

$$\mathbf{U}_{\mathbf{l}\kappa,\mathbf{l}'\kappa'}(\mathbf{R}) = -(60)^2\left[\left(\frac{\mathbf{A}}{\mathbf{R}^6}\right)\frac{1-2(\mathbf{R_B}/\mathbf{R})^2+8/3(\mathbf{R_B}/\mathbf{R})^4}{\left\{1-(2\mathbf{R_B}/\mathbf{R})^2\right\}^3} - \mathbf{B}e^{-\alpha\mathbf{R}}\left(\frac{\sinh(\alpha\mathbf{R_B})}{\alpha\mathbf{R_B}}\right)^2\left\{1+\frac{2}{\alpha\mathbf{R}}\left(1-\frac{\alpha\mathbf{R_B}}{\tanh(\alpha\mathbf{R_B})}\right)\right\}\right] \tag{3}$$

Where R is the distance between the centers of two spheres (bucky-balls), and $R_B$ is the radius of a bucky-ball.

The van der Waal's interaction between an alkali metal ion and a $C_{60}$ molecule separated by a distance r is given by

$$\mathbf{U}_{l\kappa,l'\kappa'} = -\frac{60\mathbf{A}}{\mathbf{r}^6}\left[\frac{\left(1+\left(\frac{\mathbf{R_B}}{\mathbf{r}}\right)^2\right)}{\left(1-\left(\frac{\mathbf{R_B}}{\mathbf{r}}\right)^2\right)^4}\right] + \frac{60.\mathbf{B}.\mathbf{e}^{-\alpha r}}{\mathbf{R_B}\alpha}\left[-\frac{\mathbf{R_B}}{\mathbf{r}}\cosh(\mathbf{R_B}\alpha) + \left(1+\frac{1}{\mathbf{r}\alpha}\right)\sinh(\mathbf{R_B}\alpha)\right] \tag{4}$$

**Free Rotation Model for Doped Solid**

Among the doped $A_nC_{60}$ solids the structure changes from fcc to bcc as n increases. For n =1to 3 the solids are fcc and for n=4 and 6 they are bct and bcc. For fcc structure (that of pure $C_{60}$ solid), available voids are four octahedral sites and eight tetrahedral sites per unit conventional cell. In $A_1C_{60}$, the $C_{60}$ molecules occupy fcc positions and dopants occupy octahedral voids or alternate tetrahedral voids. The $A_3C_{60}$ system is formed by filling all twelve voids. In $A_4C_{60}$ and $A_6C_{60}$ solids, $C_{60}$ molecules occupy bcc sites and alkali atoms sit on (1.5,.25) positions. In $A_6C_{60}$ all such twelve sites are filled whereas in $A_4C_{60}$ only eight of these sites are filled such that four dopant sites about any of the axes are not filled. These known structures enter as input to our calculations.

To study the charge state of the $C_{60}$ molecule we assume that some of the charge is localised on shell and the remaining electrons form a nearly free electron gas, which screens the Coulomb interactions between various ions. For a monovalent metal dopant such as K, n electrons per $C_{60}$ molecule are released to form $(K^+)_nC_{60}^{(n-x)-}$ which is partly an ionic solid. From each $C_{60}^{n-}$ ball, x electrons are released to form the electron gas while n-x electrons are localised on it. The electron density is determined as 4x and 2x electron per unit cell in the available free space in volume $a^3$,i.e., excluding the volume of the $C_{60}$ molecules for fcc structure (n=1and 3) and for bcc (n=4 and 6) respectively.

The screening length λ is taken to be that of a free electron gas, i.e.,

$$\lambda = \left( 6\pi n_0 e^2 / \varepsilon_F \right)^{-1/2} \tag{5}$$

where $n_0$ is the electron density and $\varepsilon_F$ is free-electron Fermi energy. This gives $\lambda^{-1} \cong 2.73x10^8 n_0^{1/6}$, where $n_0$ is the electron density.

The screened Coulomb potential between any two ions in the system can be symbolically represented as:

$$U^{sc}_{l\kappa,l'\kappa'} \equiv U^{sc}_{C_{60}^- - C_{60}^-}, U^{sc}_{C_{60}^- - A^+}, U^{sc}_{A^+ - A^+} \tag{6}$$

depending on the types of ions at sites **lκ** and **l'κ'.** These terms depend on the distance R between them as:

$$\mathbf{U}^{\mathbf{sc}}_{\mathbf{C_{60}^- - C_{60}^-}} = \frac{\mathbf{e}^2(3-\mathbf{x})^2}{\mathbf{R}} \mathbf{e}^{-\mathbf{R}/\lambda} \{\sinh(\mathbf{R_B}/\lambda)/(\mathbf{R_B}/\lambda)\}^2 \tag{7}$$

$$\mathbf{U}^{\mathbf{sc}}_{\mathbf{C_{60}^- - A^+}} = -\frac{\mathbf{e}^2(3-\mathbf{x})}{\mathbf{R}} \mathbf{e}^{-\mathbf{R}/\lambda} \{\sinh(\mathbf{R_B}/\lambda)/(\mathbf{R_B}/\lambda)\} \tag{8}$$

$$U^{sc}_{A^+ - A^+} = (e^2/R)\exp(-R/\lambda) \tag{9}$$

**Orientation Model for Doped $C_{60}$ Solids**

For $A_nC_{60}$ solid we assume that (n-x)/60 electrons are localised on each carbon atom of $C_{60}$ molecule and x electrons are forming electron gas which screens the coulomb interaction among ions. To avoid unnecessary computation we have compared free rotation model $C_{60}$-$C_{60}$ interaction with orientation. It is observed that for $C_{60}$ molecules free rotation model is a good approximation for distances greater than nine $A^0$. However, for $C_{60}$-$C_{60}$ interaction, which includes screened coulomb interaction also, we sum over all 60 carbon atoms of one molecule to all 60 carbon atoms of another molecule.

$$U_{l\kappa,l'\kappa'} = \sum_{ij} V(r_{ij}) \tag{10}$$

$$\mathbf{V(r)} = -\mathbf{A}/\mathbf{r}^6 + \mathbf{B}\ \exp(-\alpha\mathbf{r}) + 331.7972\,\frac{((\mathbf{n}-\mathbf{x})/60)^2}{\mathbf{r}}\mathbf{EXP}\left(-\mathbf{r}/\lambda\right) \tag{11}$$

For $C_{60}$-A interaction ( Van der Waal and screened coulomb), we sum over all 60 carbon atoms of $C_{60}$ molecule to A.

$$\mathbf{U}_{\mathbf{l}\kappa,\mathbf{l}'\kappa'} = \sum_{\mathbf{i}} \mathbf{V}(\mathbf{r_i}) \tag{12}$$

$$\mathbf{V(r)} = -\mathbf{A}/\mathbf{r}^6 + \mathbf{B}\exp(-\alpha\mathbf{r}) - 331.7972\,.\frac{((\mathbf{n}-\mathbf{x})/60)}{\mathbf{r}}\mathbf{EXP}\left(-\mathbf{r}/\lambda\right) \tag{13}$$

For A-A interaction ( Van der Waal and screened Coulomb)

$$\mathbf{U}_{\mathbf{lk,l'k'}} = -\mathbf{A}/\mathbf{r}^6 + \mathbf{B}\exp(-\alpha\mathbf{r}) + \frac{331.7972}{\mathbf{r}}\mathbf{EXP}\left(-\mathbf{r}/\lambda\right) \tag{14}$$

## Lattice Sum

The intermolecular contribution to total potential energy $\Phi$ can be obtained by carrying out the lattice sums, knowing the position of the lattice points,

$$\Phi_{\mathbf{mol-mol}} = \tfrac{1}{2} \sum_{\mathbf{l}\kappa,\mathbf{l}'\kappa'}{}' \mathbf{U}_{\mathbf{l}\kappa,\mathbf{l}'\kappa'} \tag{15}$$

The calculation presented here however uses the freely rotating model (eq.3). For the pure $C_{60}$ system this gives sufficiently accurate bulk properties.

## Energies of Anion and Cation in the Solid

The electron-affinity of $C_{60}$ is $E_A$=2.6 eV [4], and the on-shell Coulomb repulsion is U≈ 1.3 eV [5]. This gives the energy (in eV) required to add m electrons on a $C_{60}$ shell as

$E_m$ = -2.6 m + 0.65 m (m-1). (16)

In other words the energy per additional electron in HOMO is

$\varepsilon$ = -2.6 +.65 (m-1) (17)

Let $E_I$ denote the ionisation energy of the alkali atom ($E_I$ = 4eV for K). This means the s-level of A-atom is $E_I$ eV below zero. The position of the LUMO band is thus given by (compare eq. 11) -2.6 + .65 (n-x-1). Thus, we express the total potential energy of a monovalent -atom -doped $C_{60}$ solid in the following form.

$$\Phi = \frac{1}{2} \sum_{lk,l'\kappa'}' \left[ U^{vdW}_{l\kappa,l'\kappa'} + U^{sc}_{l\kappa,l'\kappa'} \right] - E_I x + \{-2.6 + .65((n-x)-1)\}(n-x) + nE_I. \quad (18)$$

Thus the total energy is expressed as a function of parameters a (lattice constant) and x, and has to be minimised w.r.t. both in order to obtain equilibrium i.e. minimum cohesive energy.

**Bulk Modulus**

Consequently minimization of Cohesive Energy yield equilibrium lattice constant $a_0$ and x (fraction of electron forming gas). Plot of cohesive energy versus volume at constant x (equilibrium value of x) is used to find second derivative of cohesive energy w.r.t. V i.e. U''. Bulk modulus is given by

$$\mathbf{B} = \left[ \mathbf{V} \frac{\partial^2 \mathbf{U}}{\partial \mathbf{V}^2} \right]_{\mathbf{V}=\mathbf{V}_0} \quad (19)$$

## RESULTS

We present here calculations for $K_1C_{60}$, $K_3C_{60}$, $K_4C_{60}$ and $K_6C_{60}$ . We have calculated the equilibrium fraction of electronic charge transferred to $C_{60}$ molecule (n-x), lattice constant ($a_0$), total cohesive energy (CE) and bulk modulus (B). Our model calculations shows complete charge transfer to $C_{60}$ molecule in all $K_nC_{60}$ systems. We have calculated lattice constant ($a_0$) and total cohesive energy (CE) for $K_2C_{60}$ system also which does not exist. Since the sum of cohesive energies of $K_1C_{60}$ and $K_3C_{60}$ exceeds twice the cohesive energy of $K_2C_{60}$ system. So Cohesive energy favours the formation of $K_1C_{60}$ and $K_3C_{60}$. We have compared our results with Band calculations[6] and some experimental values.

## Table II

| $K_xC_{60}$ (X) | Lattice Constant ($A^0$) Calc. | Lattice Constant ($A^0$) Exp. | Cohesive Energy (Kcal/Mol) Calc. | Cohesive Energy (Kcal/Mol) Band Cal. | Bulk Modulus (GPa) Calc. | Bulk Modulus (GPa) Exp. |
|---|---|---|---|---|---|---|
| 1 | 14.02 | 14.07[8] | 90 | 184 | 24 | ---- |
| 2* | 14.27 | ---- | 195 | 449 | ---- | ---- |
| 3 | 14.30 | 14.24[9] | 348 | 552 | 27 | 28[6] |
| 4 | (a) 11.5<br>(c) 10.8 | 11..9[10]<br>10.8[10] | 508 | ----- | 51 | ---- |
| 6 | 11.25 | 11.39[11] | 997 | ----- | 75 | ---- |

* No Experimental Data is available for $K_2C_{60}$ as it is not a stable system.

Lattice constant is in good agreement with experimental results in all $K_nC_{60}$ systems. The $C_{60}$ molecule has great tendency to accept six electrons to the most in $A_nC_{60}$ system. Hence, for n=1 to 6 ionic type calculations give reasonably good results. Orientation has least effect on absolute value of cohesive energy. So free rotation model is reasonably a good approximation.

**Acknowledgement**

This work was supported by Department of Science and Technology, Government of India by Research grant no. SP/S2/M13/96, on "Phonon Dynamics of Fullerenes and Derivatives". RK gratefully acknowledges support for this work by University Grant Commission, New Delhi, India.

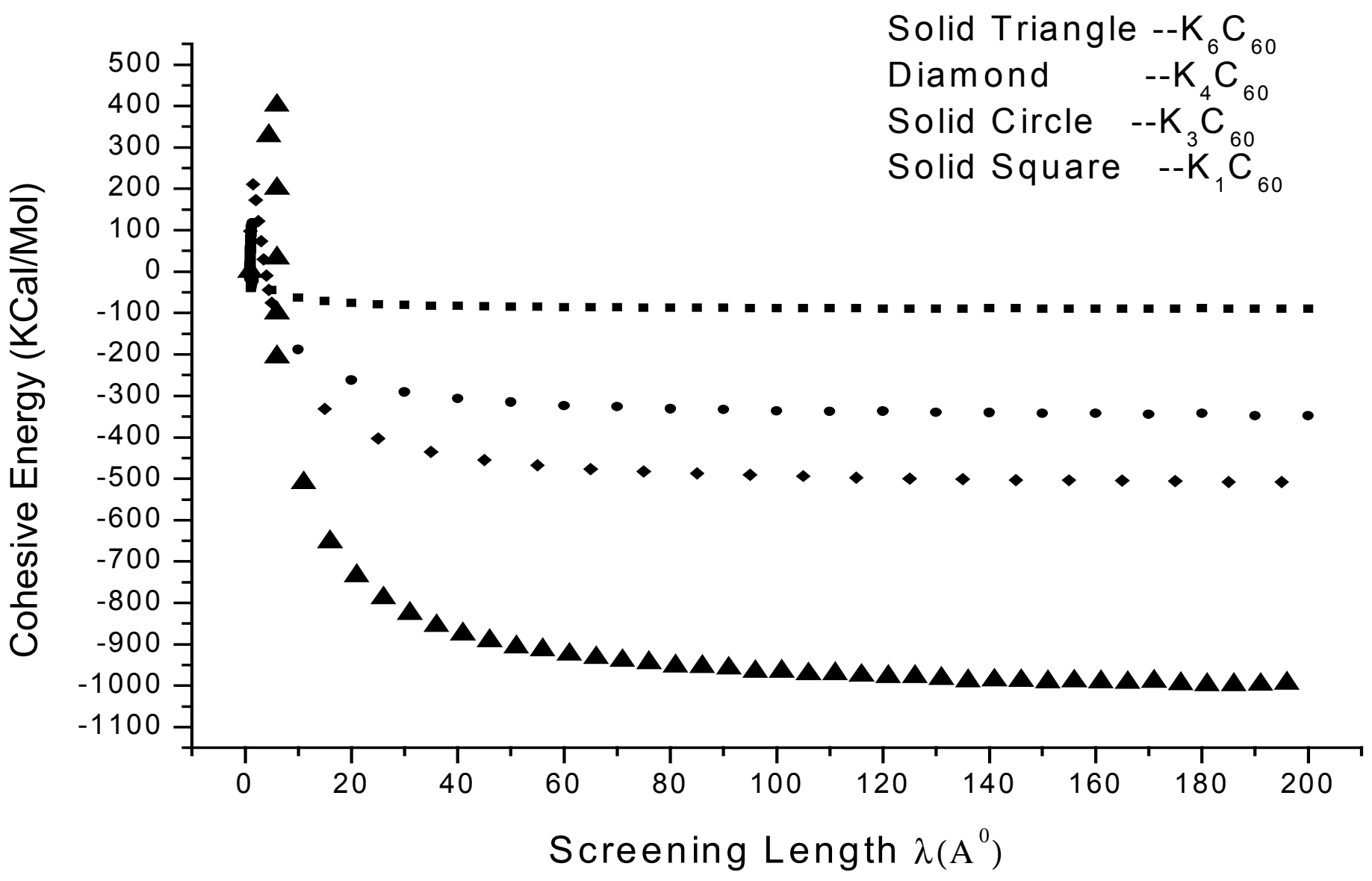

Fig.1 Calculated cohesive energy for $K_nC_{60}$ systems as a function of screening length.

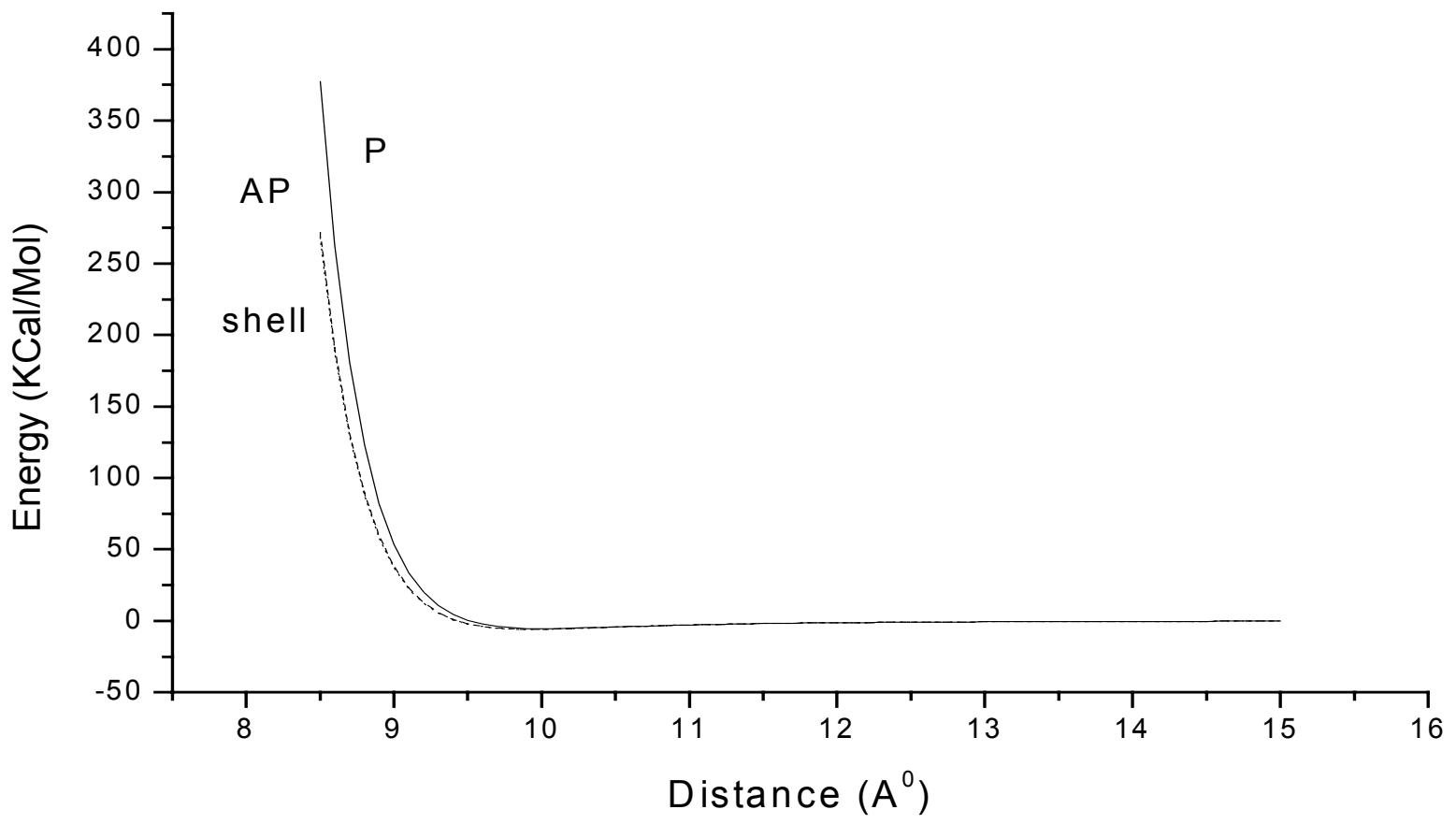

Fig.2 Calculated energy (orientation and shell approximation) between two $C_{60}$ molecules as a function of Distance. P and AP correspond to parallel and antiparallel orientations of two $C_{60}$ molecules where as Shell corresponds to shell approximation.

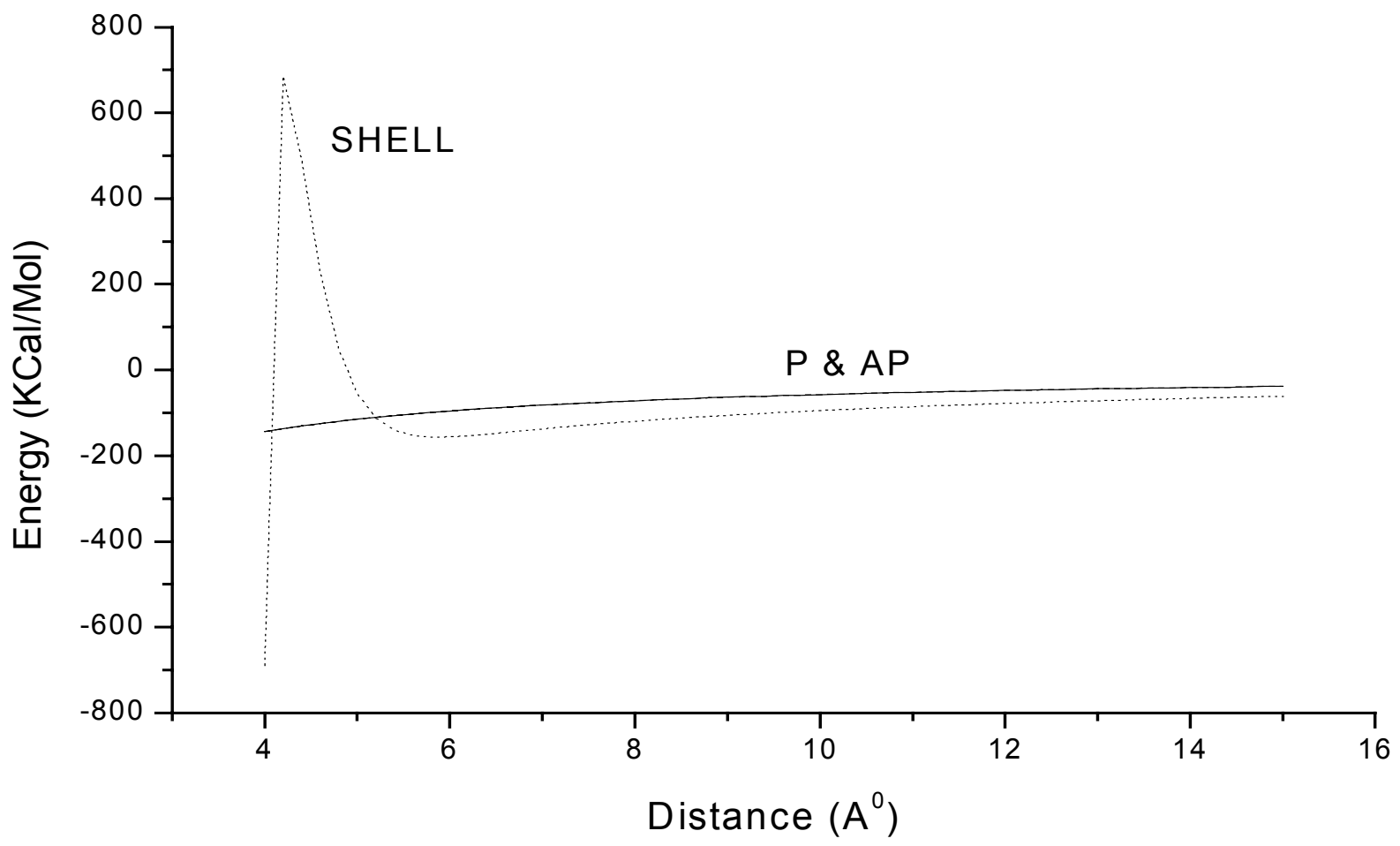

Fig 3 Calculated Total energy for $C_{60}$-K (tetrahedral) as a function of distance between K and $C_{60}$ molecule. P and AP correspond to parallel and antiparallel orientations of two $C_{60}$ molecules where as Shell corresponds to shell approximation.

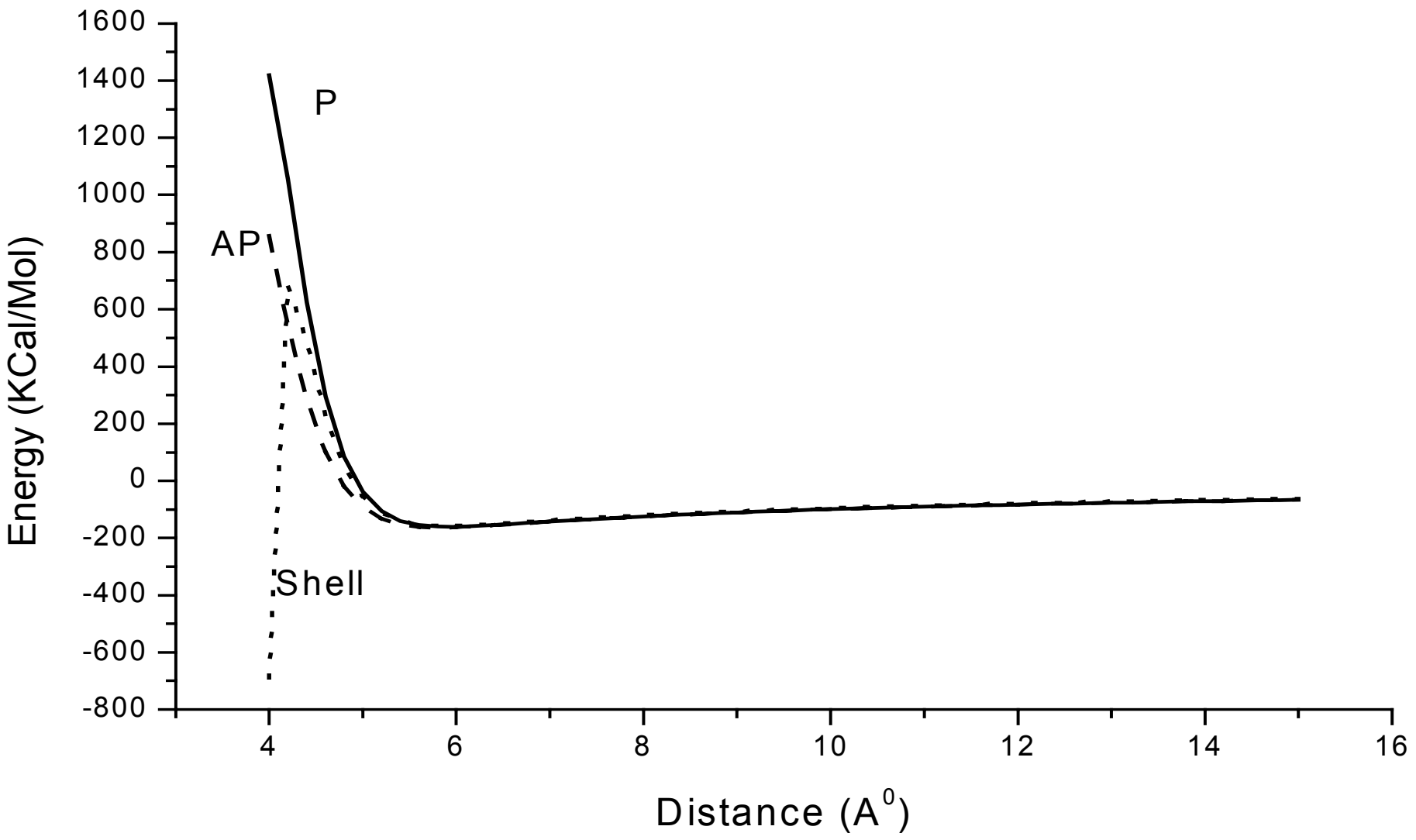

Fig 4 Calculated Total energy for $C_{60}$-K (octahedral) as a function of distance between K and $C_{60}$ molecule. P and AP correspond to parallel and antiparallel orientations of two $C_{60}$ molecules where as Shell corresponds to shell approximation.